# Reply to the Comment by B. Andresen

SUMIYOSHI ABE [1, 2, 3 (a)]

[1] *Department of Physical Engineering, Mie University, Mie 514-8507, Japan*

[2] *Institut Supérieur des Matériaux et Mécaniques Avancés, 44 F. A. Bartholdi,*
   *72000 Le Mans, France*

[3] *Inspire Institute Inc., Alexandria, Virginia 22303, USA*

**Abstract**    All the comments made by Andresen's comments are replied and are shown not to be pertinent. The original discussions [ABE S., *Europhys. Lett.* **90** (2010) 50004] about the absence of nonextensive statistical mechanics with $q$-entropies for classical continuous systems are reinforced.

_______________________________________

[(a)] E-mail: suabe@sf6.so-net.ne.jp

In his article [1], Andresen makes five comments on Ref. [2], which are listed as follows.

1) One cannot axiomatically define what is physical, since whether physical or not is a matter of taste.

2) The natural constants appear in classical physics. The presence of Planck's constant in classical thermodynamics is not problematic.

3) If one considers deeper, the dependence of the theory on Planck's constant is the crux of why it is necessary to define nonextensive entropies.

4) The fact that the total entropy is not a sum of subentropies is not a violation of physical behavior.

5) Defining the measure as in Ref. [2] with a uniform limit $n \to \infty$ can be modified in order to realize the continuous limit of $q$-entropies [3].

Below, we examine all of these comments.

First, let us ask the following question regarding 2). Among the natural constants, $c$ (velocity of light), $e$ (elementary charge), $G$ (gravitational constant) and $h$ (Planck's constant), which is relevant to the foundations of thermostatistics? The answer is ... $h$! Without $h$, it is not possible to calculate the entropy in the $\Gamma$-space, whereas one can always consider a statistical-mechanical system of nonrelativistic electrically-neutral particles with negligible gravitational interaction.

Regarding 3), there is a simple way to see why it is wrong. The virial theorem [4,5] tells us about the long-time average (and therefore is more fundamental than the classical ensemble average). This theorem states that

$$PV = \frac{2}{3}\overline{E} - \frac{1}{6}\overline{\sum_{i,j} r_{ij}\frac{d\phi(r_{ij})}{dr_{ij}}}. \qquad (1)$$

Here, $P$, $V$ and $\overline{E}$ are the pressure, volume and average kinetic energy. $\phi(r_{ij})$ is a two-body potential between the $i$th and $j$th particles with distance $r_{ij} = |\mathbf{r}_i - \mathbf{r}_j|$, where $\mathbf{r}_i$ ($\mathbf{r}_j$) denotes the position of the $i$th ($j$th) particle. The over-bar denotes the long-time average. Eq. (1) is a generalization of Bernoulli's formula. The second term on the right-hand side of this equation is approximated to be

$$-\frac{4\pi N^2}{6V}\int_0^R dr\ r^3\frac{d\phi(r)}{dr}, \qquad (2)$$

where $N$ and $R = (3V/4\pi)^{1/3}$ are the number of the particles and the system size, respectively. Therefore, even for a long-range interaction (such as $\phi(r) \sim 1/r$), with which the system is *nonextensive*, there is no room for Planck's constant to appear in classical thermodynamics, as usual. On the other hand, it is found [6] that, in classical nonextensive statistical mechanics [3], various thermodynamics equations and quantities such as the equation of state and specific heat depend explicitly on Planck's constant, leading to a serious conceptual difficulty.

The comment 4) might look legitimate at first glance. However, this statement is actually obscure. What is importance is correlation. If subsystems are not independent, then the total entropy has to be given in terms of the conditional entropies as well as the marginal entropies [7,8].

Next, let us examine 5). Clearly, Andresen fails to present any examples of such measure. Therefore, here we employ the work in Ref. [9] as an explicit example. The authors of Ref. [9] try to define a measure, with which a divergence in the limit $n \to \infty$ might become an additive constant for $q$-entropies. Unfortunately, they are obviously unsuccessful to do so. The measure they construct depends explicitly on the probability distribution itself. Accordingly, one cannot calculate even the normalization of the distribution. It should be noted that the integral employed in calculating the $q$-Gaussian distributions etc. in nonextensive statistical mechanic is based on the ordinary measure and not on a pathological one.

Finally, we make remarks on the comment 1). In Ref. [2], the word "physical" is used for the physical entropy, which is the equilibrium/nonequilibrium thermodynamic entropy. There is nothing metaphysical there. Also, it has been well-known at least since Hilbert [10] that one cannot talk about physics in an axiomatic manner.

In conclusion, the comments made by Andresen's comments are fully replied and refuted. The original discussions made in Ref. [2] about the absence of nonextensive statistical mechanics with $q$-entropies for classical continuous systems are reinforced. We also note that the questions developed in Ref [2] were recently examined in Ref. [11], demonstrating from a different viewpoint the limits of validity of nonextensive statistical mechanics with $q$-entropies for continuous systems.

\* \* \*

This work was supported in part by a Grant-in-Aid for Scientific Research from the Japan Society for the Promotion of Science (JSPS).